\documentclass[conference]{IEEEtran}
\IEEEoverridecommandlockouts
\usepackage{cite}
\usepackage{amsmath,amssymb,amsfonts}
\usepackage{algorithmic}
\usepackage{graphicx}
\usepackage{textcomp}
\usepackage{xcolor}
\def\BibTeX{{\rm B\kern-.05em{\sc i\kern-.025em b}\kern-.08em
    T\kern-.1667em\lower.7ex\hbox{E}\kern-.125emX}}
\usepackage{amsmath,amssymb}
\usepackage{tikz,ifthen,calc,graphicx,pgfplots}%
\usepackage{balance}
\usepackage[keeplastbox]{flushend}

\def\A{{\mathbf A}}

\def\F{{\mathbf F}}
\def\H{{\mathbf H}}
\def\I{{\mathbf I}}
\def\L{{\mathbf L}}
\def\Q{{\mathbf Q}}
\def\R{{\mathbf R}}
\def\S{{\mathbf S}}
\def\T{{\mathbf T}}

\def\s{{\mathbf s}}

\def\u{{\mathbf u}}
\def\v{{\mathbf v}}
\def\w{{\mathbf w}}
\def\x{{\mathbf x}}
\def\y{{\mathbf y}}
\def\z{{\mathbf z}}

\def\0{{\mathbf 0}}

\def\E{ \operatorname{E}}
\def\diag{ \operatorname{diag}}

\usetikzlibrary{external}
\begin{document}

\title{Two-Channel Passive Detection Exploiting Cyclostationarity}

\author{\thanks{The work of S. Horstmann and P. J. Schreier was supported by the German Research Foundation (DFG) under grant SCHR 1384/6-1. The work of D. Ram{\'\i}rez was supported by the Ministerio de Ciencia, Innovaci{\'o}n y Universidades under grant TEC2017-92552-EXP (aMBITION), by the Ministerio de Ciencia, Innovaci{\'o}n y Universidades, jointly with the European Commission (ERDF), under grants TEC2015-69868-C2-1-R (ADVENTURE) and TEC2017-86921-C2-2-R (CAIMAN), by The Comunidad de Madrid under grant Y2018/TCS-4705 (PRACTICO-CM), and by the German Research Foundation (DFG) under grant RA 2662/2-1.}
\IEEEauthorblockN{Stefanie~Horstmann}
\IEEEauthorblockA{\textit{Signal and System Theory Group} \\
\textit{University of Paderborn}\\
Paderborn, Germany \\
stefanie.horstmann@sst.upb.de}
\and
\IEEEauthorblockN{David~Ram\'irez}
\IEEEauthorblockA{\textit{Department of Signal Theory and Communications} \\ 
\textit{Universidad Carlos III de Madrid}\\
\textit{Gregorio Mara{\~n}{\'o}n Health Research Institute}\\
Legan{\'e}s, Spain \\
david.ramirez@uc3m.es}
\and
\IEEEauthorblockN{Peter~J.~Schreier}
\IEEEauthorblockA{\textit{Signal and System Theory Group} \\
\textit{University of Paderborn}\\
Paderborn, Germany \\
peter.schreier@sst.upb.de}}

\maketitle

\begin{abstract}
This paper addresses a two-channel passive detection problem exploiting cyclostationarity. Given a reference channel (RC) and a surveillance channel (SC), the goal is to detect a target echo present at the surveillance array transmitted by an illuminator of opportunity equipped with multiple antennas. Since common transmission signals are cyclostationary, we exploit this information at the detector. Specifically, we derive an asymptotic generalized likelihood ratio test (GLRT) to detect the presence of a cyclostationary signal at the SC given observations from RC and SC. This detector tests for different covariance structures. Simulation results show good performance of the proposed detector compared to competing techniques that do not exploit cyclostationarity.
\end{abstract}


\section{Introduction}

We consider a two-channel multiple-input multiple-output (MIMO) passive detection problem motivated by a passive radar application. Specifically, we consider a passive bistatic radar consisting of one receiver and one transmitter each equipped with multiple antennas. However, the transmitter is non-cooperative and also referred to as illuminator of opportunity (IO). That is, the illuminator operates independently of the passive radar system and would typically be a commercial broadcast system such as DVB-T or, for instance, navigation satellites \cite{Griffiths2005}. In order to detect a moving object a two-channel passive radar system consists of a reference channel (RC) and a surveillance channel (SC). The reference array observes a noisy version of the transmitted signal from the IO, whereas the surveillance array there is the reflected signal from the target. If there is no target present, only noise is measured at the SC. Direct-path signals from the IO to the SC are assumed to be canceled by, for instance, directional antennas.

Our goal is to detect whether there is a target echo at the SC, i.e., whether there is correlation between the signals observed at the RC and the SC.
A common approach to solve this detection problem is based on the cross-correlation of the signals at SC and RC. However, this approach is only suboptimal due to noise at the RC \cite{Liu2015}. Generalized likelihood ratio tests (GLRT) were derived in \cite{Santamaria2016, Wang2016, Santamaria20172, Santamaria2017} for various assumptions on the signal and noise models and considering unknown stochastic waveforms. The authors in \cite{Hack2014} and \cite{Howard2013} derived the GLRT for unknown deterministic waveforms in temporally and spatially white noise. Furthermore, \cite{Howard2013} and \cite{Howard2016} also presents Bayesian tests for the same problem.

These detectors typically assume that the transmitted signals are temporally white. However, digital communication signals as transmitted by potential IOs are cyclostationary (CS) processes \cite{Gardner1987}. This property was exploited, for instance, in \cite{Gardner1993} and \cite{Gelli1996}, which derived locally optimum tests for detection with a single array for known signal statistics and different assumptions on the noise (temporally and spatially white Gaussian  in \cite{Gardner1993} and non-Gaussian in \cite{Gelli1996}). The authors in \cite{Ramirez2015} derived the GLRT and locally most powerful invariant test for the single array case for unknown waveforms in temporally colored and spatially correlated noise, which was specialized to various noise structures in \cite{Pries2016, Pries2018}. 

In this paper, we derive the GLRT for the two-channel passive detection problem that aims at detecting the presence of cyclostationarity in the SC given the additional (reference) channel. Since the derivation requires the estimation of covariance matrices with block-Toeplitz structure, we make use of an asymptotic result from \cite{Ramirez2015}, which allows us to find approximate closed-form estimates under both hypotheses.

The paper is organized as follows: The detection problem is formulated in Section 2 followed by the derivation of the GLRT in Section 3.  In Section 4 we evaluate the performance of the detector with numerical simulations.

\section{Problem formulation}
\label{sec:problem formulation}
We consider a passive radar system consisting of an RC and an SC. Without loss of generality, we assume that both RC and SC are equipped with $L$ antennas each, but the derivations in this paper can be easily extendend to different numbers of antennas at both arrays. At the RC a noisy version of the transmitted signal by the IO is received, whereas at the SC the target echo is observed, which we assume to be synchronized in time delay and Doppler shift \cite{Santamaria2017, Zhao2017}. Furthermore, it is assumed that there is no direct-path interference present at the SC, which is a reasonable assumption considering that directional antennas are used or spatial filtering is applied. This two-channel detection problem has the two hypotheses
\begin{equation}
\begin{array}{rl}
\mathcal{H}_0: \; &\begin{cases}
\u_s[n] =  \v_s[n], \\
\u_r[n] =  \H_r[n] \ast \s[n] + \v_r[n],
\end{cases}\\
\mathcal{H}_1: \; &\begin{cases}
\u_s[n] =  \H_s[n] \ast \s[n] + \v_s[n], \\
\u_r[n] =  \H_r[n] \ast \s[n] + \v_r[n],
\end{cases}
\end{array}
\end{equation}
where $\H_s[n]\in \mathbb{C}^{L\times \rho}$ and $\H_r[n]\in \mathbb{C}^{L\times \rho}$ represent the frequency-selective channels from the IO to the reference and surveillance arrays, respectively. The additive noise terms $\v_s[n] \in \mathbb{C}^{L}$ and $\v_r[n] \in \mathbb{C}^{L}$ are assumed to be wide-sense stationary (WSS) with arbitrary temporal and spatial correlation, but the noise terms are assumed to be uncorrelated between reference and surveillance arrays. The signal $\s[n] \in \mathbb{C}^\rho$ transmitted by an IO equipped with $\rho$ antennas is assumed to be a discrete-time zero-mean second-order CS signal with cycle period $P$, i.e., its matrix-valued covariance sequence $\R_{\s\s}[n,m] = \E[\s[n]\s^H[n-m]] =  \R_{\s\s}[n+P,m]$ is periodic in $n$ with period $P$. This implies that the signal $\u_r[n]$ is a multivariate CS process with cycle period $P$ under both hypotheses, whereas $\u_s[n]$ is WSS under $\mathcal{H}_0$ and CS with cycle period $P$ under $\mathcal{H}_1$.  We assume that the cycle period $P$ is known a priori. This is a reasonable assumption since the cycle period is related to signal features such as symbol rate, carrier frequency, or cyclic prefix length, which are known if the standard used by the IO is known. If this is not the case, the cycle period may be estimated with techniques presented in, e.g. \cite{Ramirez2014, Dandawate1994}. Furthermore, we assume that $\rho \geq L$ to ensure that the covariance functions of $ \H_s[n] \ast \s[n]$ and $ \H_r[n] \ast \s[n]$ are full rank as the additional structure imposed by low-rank covariance matrices would not be exploited in this work.

It is shown in \cite{Gladyshev1961} that the vector 
\begin{equation}
\x[n]=\left[{\mathbf u}^T[nP] \; \cdots \; {\mathbf u}^T[(n+1)P-1]\right]^T,
\end{equation}
is WSS if ${\mathbf u}[n] \in \mathbb{C}^{L}$ is CS with cycle period $P$. Hence, the covariance matrix $\R_{\x\x}[n,m]=\E\left[\x[n]\x^H[n-m]\right] = \R_{\x\x}[m] \in \mathbb{C}^{LP\times LP}$ only depends on the time-shift. Moreover, the covariance matrix of a stack of $N$ realizations of $\x[n]$, i.e., 
$\y  =  \left[ \x^T[0] \; \cdots \; \x^T[N-1]\right]^T \in \mathbb{C}^{LNP}$,
is given by
\begin{equation}
\label{eq: covariance matrix x cyclo}
\R_{\y\y} = \E[\y\y^H] = \left[ \begin{array}{ccc} \R_{\x\x}[0] & \cdots & \R_{\x\x}[N-1] \\ 
\vdots & \ddots & \vdots\\
 \R_{\x\x}^{H}[N-1]& \cdots &  \R_{\x\x}[0]\end{array} \right],
\end{equation}
which is a block-Toeplitz matrix with block size $LP$.

Following the latter considerations, we stack $NP$ samples of $\u_r[n]$ and $\u_s[n]$ into the vectors
\begin{align}
\label{eq: vector x_r}
\y_r & =  \left[ {\mathbf u}^T_r[0] \; \cdots \; {\mathbf u}_r^T[NP-1]\right]^T \in \mathbb{C}^{LNP},\\
\label{eq: vector x_s}
\y_s &  =  \left[ {\mathbf u}^T_s[0] \; \cdots \; {\mathbf u}_s^T[NP-1]\right]^T \in \mathbb{C}^{LNP},
\end{align}
respectively. 
Under both hypotheses, the covariance matrix $\R_r = \E[\y_r\y_r^{H}]$ is a block-Toeplitz matrix with block size $LP$ as the signal $\u_r[n]$ is CS with cycle period $P$ under $\mathcal{H}_0$ and $\mathcal{H}_1$. The covariance matrix $\R_s = \E[\y_s\y_s^{H}]$ is a block-Toeplitz matrix with block size $LP$ under $\mathcal{H}_1$ since $\u_s[n] \in \mathbb{C}^L$ is CS with cycle period $P$, whereas $\R_s$ is block-Toeplitz with block size $L$ when $\u_s[n]$ is WSS.

Under the null hypothesis the signals $\y_s$ and $\y_r$ are uncorrelated. For this reason the covariance matrix of
\begin{equation}
\label{eq: w vector }
\w = \left[{\mathbf y}_s^T \; \; {\mathbf y}_r^T\right]^T \in \mathbb{C}^{2LNP}
\end{equation}
is given by
\begin{equation}
\label{eq: covariance matrix of w}
\R_0 = \E[\w\w^H|\mathcal{H}_0] = \left[ \begin{array}{cc}\R_{s} & \mathbf{0} \\ \mathbf{0} &\R_r \end{array} \right],
\end{equation}
with block-Toeplitz matrices $\R_s$ and $\R_r$ with block sizes $L$ and $LP$, respectively.

Under $\mathcal{H}_1$ the signals at SC and RC are correlated and the structure of $\R_1 =  \E[\w\w^H|\mathcal{H}_1]$ is more involved since its off-diagonal blocks are non-zero. For this reason, we permute the elements in $\w$ 
as
\begin{equation}
\label{eq: wtilde}
\tilde{\w} = \left(\L_{2NP,NP} \otimes \I_L\right)\w,
\end{equation}
where $\L_{2NP,NP}$ is the commutation matrix.\footnote{For the commutation matrix the following holds: $\operatorname{vec}\left(\A\right) = \L_{MN,N}\operatorname{vec}\left(\A^T\right)$ for an $M \times N$ matrix $\A$. Note that $\L_{MN,N}^T = \L_{MN,M}$.} This structure makes it easier to find a maximum-likelihood (ML) estimate. Now the vector $\tilde{\w}$ contains the samples $\u_s[n]$ and $\u_r[n]$ in alternating order. Since the vector $\left[\u_s[n]^T \; \; \u_r[n]^T\right]^T \in \mathbb{C}^{2L}$ is CS with cycle period $P$, the covariance matrix $\R_{\tilde{\w}\tilde{\w}} = \E[\tilde{\w}\tilde{\w}^H|\mathcal{H}_1]$ is a block-Toeplitz matrix with block size $2LP$. 
Hence, from \eqref{eq: wtilde} the covariance matrix of $\w$ under the alternative hypothesis is given by
\begin{equation}
\label{eq: covariance w alternative}
\R_1 =  \left(\L_{2NP,NP}^T \otimes \I_L\right)\R_{\tilde{\w}\tilde{\w}} \left(\L_{2NP,NP} \otimes \I_L\right).
\end{equation}

Finally, we assume $\u_s[n]$ and $\u_r[n]$ to be zero-mean proper complex Gaussian random processes, thus, the hypothesis test can be formulated as
\begin{equation}
\label{eq: hypotheses time domain}
\begin{array}{c}
\mathcal {H}_0: {\mathbf w} \sim\, \mathcal {CN}_{2LNP}({\mathbf 0},{\mathbf R}_0),\\ 
\mathcal {H}_1: {\mathbf w}\sim \, \mathcal {CN}_{2LNP}({\mathbf 0},{\mathbf R}_1).
\end{array} 
\end{equation}

\section{Derivation of the GLRT}
Since the covariance matrices $\R_0$ and $\R_1$ are unknown, we deal with a composite hypothesis test, which can commonly be  approached with a GLRT. To this end we have to find the ML estimates of the covariance matrices under both hypotheses. However, the covariance matrices are block-Toeplitz for which no closed-form solution exists \cite{Burg1982}. For this reason we make use of Theorem 1 in \cite{Ramirez2015}, where the authors showed that we can asymptotically (as $N \rightarrow\infty$) approximate the block-Toeplitz covariance matrix as a block-circulant covariance matrix. As block-circulant matrices can be diagonalized by the DFT, we can exploit this property in order to obtain covariance matrices that are easier to estimate. Specifically, we consider the following linear transformation of $\w$
\begin{equation}
\label{eq: z linear trafo of w}
\z = \left(\I_2 \otimes (\L_{NP,N} \otimes \I_L)(\F_{NP} \otimes \I_L)^{H}\right) \w,
\end{equation}
where $\F_{NP}$ is the DFT matrix of dimension $NP$. Hence, this transformation reorders the frequency components of the DFTs of $\w_s$ and $\w_r$ in a specific way. In the subsequent sections we will show that this transformation allows us to easily obtain the (asymptotic) ML estimates of the covariance matrix of $\z$ under both hypotheses.
The hypothesis test may be reformulated as
\begin{equation}
\begin{array}{c}
\mathcal {H}_0: \z \sim\, \mathcal {CN}_{2LNP}({\mathbf 0},\S_0),\\ 
\mathcal {H}_1: \z \sim\, \mathcal {CN}_{2LNP}({\mathbf 0},\S_1),
\end{array} 
\end{equation}
where $\S_0 = \E[\z\z^H|\mathcal{H}_0]$ and $\S_1 = \E[\z\z^H|\mathcal{H}_1]$ are the covariance matrices under each hypothesis. These covariance matrices have different structures as we will show in Sections \ref{subsec: MLE null hypothesis} and \ref{subsec: MLE alternative}.
Therefore, assuming there are $M$ independent and identically distributed (i.i.d.) realizations\footnote{Since in practice there are no i.i.d. observations available, we divide a long observation into $M$ windows and treat them as if they were i.i.d.} of $\z$, the generalized likelihood ratio (GLR) is given by
\begin{equation}
\label{eq: likelihood ratio}
\Lambda = \frac{p(\z_0,\cdots, \z_{M-1}; \hat{\S}_1)}{p(\z_0,\cdots, \z_{M-1}; \hat{\S}_0)}, 
\end{equation}
where $\hat{\S}_0$ and $\hat{\S}_1$ denote the ML estimates under $\mathcal{H}_0$ and $\mathcal{H}_1$, respectively, which are derived in the following paragraphs. Under the Gaussian assumption the likelihoods are given by 
\begin{align}
\label{eq: likelihoods}
p(\z_0,\cdots, \z_{M-1}; \hat{\S}_i) = &\frac{1}{\pi^{2LNPM}\det \left(\hat{\S}_i\right)^M} \nonumber \\
 & \times \exp \left\lbrace -M \operatorname{tr}\left(\Q\hat{\S}_i^{-1}\right)\right\rbrace,
\end{align}
where $\Q = \frac{1}{M}\sum_{m=0}^{M-1} \z_m\z_m^H$ is the sample covariance matrix of $\z$ and $i\in \left\lbrace 0,1\right\rbrace$ indicates whether it is the likelihood under $\mathcal{H}_0$ or $\mathcal{H}_1$. 
In order to obtain the GLR, we must now derive the ML estimates of the covariance matrices.

\subsection{ML estimate under the null hypothesis}
\label{subsec: MLE null hypothesis}
Let us first consider the covariance matrix of $\z$ under the null hypothesis, which is given by
\begin{equation}
\label{eq: covariance matrix z}
\S_0 = \E[\z\z^H|\mathcal{H}_0] = \left[ \begin{array}{cc} \S_{s} & \mathbf{0} \\ \mathbf{0} & \S_{r} \end{array} \right],
\end{equation}
where $\S_{s} \in \mathbb{C}^{LNP \times LNP}$ is a block-diagonal matrix with block size $L$ and $\S_{r} \in \mathbb{C}^{LNP \times LNP}$ is a block-diagonal matrix with block size $LP$. This can be observed by recalling that the transformation in \eqref{eq: z linear trafo of w} block-diagonalizes a block-circulant matrix. This can be easily verified considering the results from \cite{Ramirez2015}. 

The ML estimate of $\S_0$ can be obtained using results from complex-valued matrix differentiation \cite{Schreier2010} and is given by
\begin{equation}
\label{eq: MLE S0}
\hat{\S}_0 = \left[ \begin{array}{cc}  \diag_{L}\left(\Q_{s}\right) & \mathbf{0} \\ \mathbf{0} & \diag_{LP}\left(\Q_{r}\right) \end{array} \right],
\end{equation}
where $\Q_{s}$ and $\Q_{r}$ denote the north-west and south-east blocks of dimension $LNP \times LNP$ of the sample covariance matrix $\Q$, respectively. Moreover, $\diag_i(\A)$ denotes an operator that builds a block-diagonal matrix with block size $i$ from the diagonal blocks of $\A$.

\subsection{ML estimate under the alternative}
\label{subsec: MLE alternative}
Second, to find the ML estimate of $\S_1$ let us consider again the permutation $\tilde{\w}$ from \eqref{eq: wtilde}, which has a block-Toeplitz structured covariance matrix  $\R_{\tilde{\w}\tilde{\w}}$ with block size $2LP$. This enables us to find a closed-form (asymptotic) ML estimate. 

Similar to the previous section
let us consider the linear transformation of $\tilde{\w}$ given by
\begin{equation}
\label{eq: ztilde linear trafo of w tilde}
\tilde{\z} = (\L_{NP,N} \otimes \I_{2L})(\F_{NP} \otimes \I_{2L})^{H}\tilde{\w}.
\end{equation}
 Again we can asymptotically block-diagonalize $\R_{\tilde{\w}\tilde{\w}}$ by the latter transformation, i.e., $\tilde{\S}_1 = \E\left[\tilde{\z}\tilde{\z}^H|\mathcal{H}_1\right]$ is the block-diagonal matrix with block size $2LP$. 
Considering that $\z = \T\tilde{\z}$ with permutation matrix
\begin{align}
\label{eq: T matrix}
\T = \left(\L_{2NP,NP}^T \otimes \I_L\right),
\end{align}
which follows from \eqref{eq: wtilde}, \eqref{eq: z linear trafo of w}, and \eqref{eq: ztilde linear trafo of w tilde}, the ML estimate of $\tilde{\S}_1$ is given by
\begin{equation}
\label{eq: MLE S tilde}
\hat{\tilde{\S}}_1 = \diag_{2LP}\left(\tilde{\Q}\right),
\end{equation}
where $\tilde{\Q} = \T^{T}\Q \T$.
As we are interested in an estimate of $\S_1$ rather than $\hat{\tilde{\S}}_1$, we exploit the invariance property of the ML estimate to find
\begin{equation}
\label{eq: MLE S1}
\hat{\S}_1 = \T\hat{\tilde{\S}}_1 \T^T.
\end{equation}

\subsection{GLRT}
\label{subsec: GLR}
Putting the pieces together, the GLR is given by
\begin{align}
\label{eq: glr plugged in estimates final}
\Lambda^{\frac{1}{M}} &= \frac{\det\left( \hat{\S}_0\right)}{\det\left( \hat{\S}_1\right)} \\
& = \frac{\prod_{k=1}^{NP}\det\left(\left[\Q_{s}\right]^k_L\right) \prod_{l=1}^{N}\det\left(\left[\Q_{r}\right]^l_{LP}\right)}{\prod_{l=1}^{N}\det\left( \left[\tilde{\Q}\right]^l_{2LP}\right)}, 
\end{align}
where $\left[\A\right]^k_K$ denotes the $k$th diagonal block of size $K$ of matrix $\A$, and we exploited properties of the determinant of block-diagonal matrices and that $\T$ is an orthogonal matrix, i.e., \mbox{$(\det(\T))^2 = 1$}.
Finally, the GLRT is
\begin{equation}
\label{eq: glrt}
\Lambda  ^{\frac{1}{M}}  \overset{\mathcal{H}_1}{\underset{\mathcal{H}_0}{\gtrless}}  \eta.
\end{equation}

We now need to find a threshold $\eta$ that assures a given probability of false alarm. 
To this end it should be noted that the test statistic is invariant to a multiplication with any matrix of the structure of $\S_0$ in \eqref{eq: covariance matrix z}.
In the time-domain this is equivalent to an invariance to filtering, i.e., circular convolution of $\u_s[n]$ and circular convolution of $\x_r[n]$. Hence, we can use numerical simulations with a white process under the null hypothesis to obtain the threshold, which can be applied for any arbitrary covariance matrix $\S_0$ (asymptotically).


\section{Numerical results}
\label{sec:numerical results}

In this section we evaluate the performance of the proposed detector using Monte Carlo simulations. According to our model we generate a CS signal $\s[n]$ as a QPSK-signal with raised-cosine pulse shaping and roll-off factor $1$. The symbol rate is $R_s=600$\ Kbauds. Together with a sampling frequency $f_s = 1.2$\ MHz this yields a cycle period of $P=2$. The frequency-selective channels $\H_r[n]$ and $\H_s[n]$ are both Rayleigh-fading channels with a delay spread of 10 times the symbol duration. Moreover, we draw a new channel realization in every Monte Carlo simulation. The independent noises at SC and RC are both colored Gaussian generated with a moving average filter of order $20$ and correlated among antennas.

The benchmark techniques are the correlated subspace detector proposed in \cite{Santamaria2017} and the popular cross-correlation detector \cite{Griffiths2005,Liu2015}. It should be noted that the cross-correlation detector does not require any prior knowledge, whereas the correlated subspace detector needs to know the number of antennas $\rho$ at the IO and the proposed technique also needs to know the cycle period $P$.

To evaluate the performance of the proposed GLRT we choose a scenario with $\rho=2$  transmit antennas at the IO  and $L=2$ receive antennas at both the SC and RC. Furthermore, we choose $N=32$ and $M=16$, i.e., we generate a sequence of length $NM$, which we cut into $M$ pieces of length $N$. The particular choice of $N$ and $M$ is a bias-variance trade-off. As it was shown in \cite{Ramirez2010} for a similar problem, if $NM$ is small, it is beneficial to sacrifice some spectral resolution (smaller $N$) in order to decrease the variance of the estimate (larger $M$). If there are more samples available, they can be used to achieve a better spectral resolution (larger $N$).

\begin{figure}
\begin{center}
\includegraphics[scale=1]{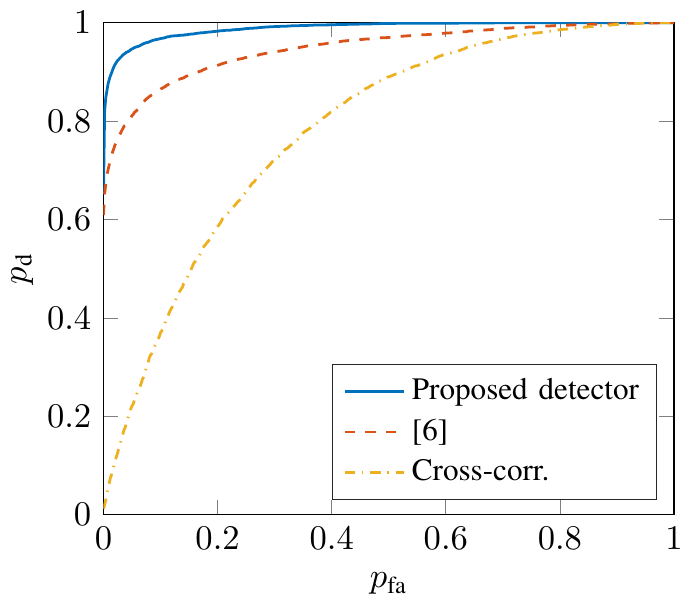}
\end{center}
\vspace{-0.25cm}
\caption{ROC for a scenario with $P=2$, $N=32$, $M=16$, $L=\rho=2$, $\text{SNR}_s = -10$\ dB and $\text{SNR}_r = 0$\ dB.}
\label{fig: roc}
\vspace{-0.25cm}
\end{figure}

\begin{figure}
\begin{center}
\includegraphics[scale=1]{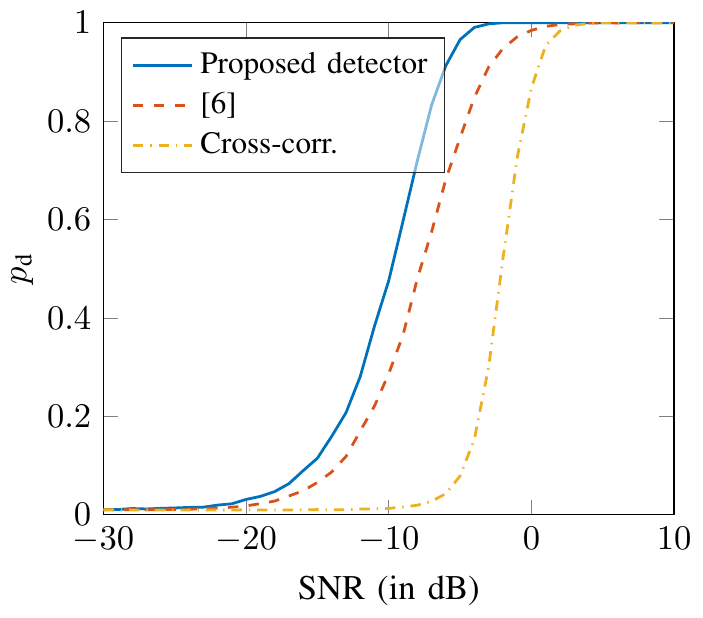}
\end{center}
\vspace{-0.25cm}
\caption{Probability of detection vs. $\text{SNR}$ (for $\text{SNR}_s=\text{SNR}_r$) for a scenario with $P=2$, $N=32$, $M=16$, $L=\rho=2$ and $p_{\text{fa}}= 0.01$.}
\label{fig: pd vs snr}
\vspace{-0.25cm}
\end{figure}

Figure \ref{fig: roc} shows the receiver operating characteristic (ROC) curve for fixed $\text{SNR}_s=-15$\ dB and $\text{SNR}_r=0$\ dB at the SC and RC, respectively. As can be seen the proposed GLRT outperforms both the technique from \cite{Santamaria2017} and the cross-correlation detector. Moreover, in Figure \ref{fig: pd vs snr} we show the probability of detection $p_{\text{d}}$ versus the SNR, where we assume that $\text{SNR}_s=\text{SNR}_r$ and probability of false alarm $p_{\text{fa}}= 1\%$. Again the GLRT performs better than the detector from \cite{Santamaria2017} and the cross-correlation detector.

\section{Conclusion}
\label{sec: conclusion}
We have derived the GLRT for a two-channel passive detection problem for cyclostationary processes. The proposed technique tests for different covariance structures under the null hypothesis and alternative. Its main advantage is that it exploits the fact that digital communication signals are cyclostationary. The simulation results show that the proposed technique outperforms the benchmark detectors, which do not use this fact.

\end{document}